\documentclass[aps,pra,twocolumn,showpacs]{revtex4-1} 

\usepackage{graphicx} 
\usepackage{amsmath}
\usepackage{bm}    
\usepackage{amssymb}  

\newcommand{\ket}[1]{\left|#1\right\rangle}
\newcommand{\bra}[1]{\left\langle#1\right|}
\newcommand{\braket}[2]{\left\langle#1|#2\right\rangle}

\begin{document}

\title{Non-Markovian evolution of photonic quantum states in atmospheric turbulence}

\author{Filippus S. Roux}
\email{fsroux@csir.co.za}
\affiliation{CSIR National Laser Centre, PO Box 395, Pretoria 0001, South Africa}
\affiliation{School of Physics, University of Witwatersrand, Johannesburg 2000, South Africa}

\begin{abstract}
The evolution of the spatial degrees of freedom of a photon propagating through atmospheric turbulence is treated as a non-Markovian process. Here, we derive and solve the evolution equation for this process. The turbulent medium is modeled by a sequence of multiple phase screens for general turbulence conditions. The non-Markovian perspective leads to a second-order differential equation with respect to the propagation distance. The solution for this differential equation is obtained with the aid of a perturbative analysis, assuming the turbulence is relatively weak. We also provide another solution for more general turbulence strength, but where we introduced a simplification to the differential equation.
\end{abstract}


\maketitle

\section{Introduction}

The scintillation that a photonic quantum state experience as it propagates through a turbulent atmosphere is a topic of considerable importance for free-space quantum communication. The evolution of the quantum state in this scenario can be considered, using a single phase screen (SPS) model \cite{paterson}, provided that the scintillation remains weak. Although the SPS model is used in most of the work that has been done in this field \cite{sr,qturb4,qturb3,pors,malik,toddbrun,leonhard}, a more accurate multiple phase screen (MPS) approach has been proposed recently \cite{ipe,iperr,lindb,notrunc}. The MPS approach is based on the principle of infinitesimal propagation, which allows one to derive an equation for the evolution of the quantum state, called the infinitesimal propagation equation (IPE). The IPE is a first-order differential equation with respect to the propagation distance, which can be solved \cite{notrunc} to obtain an expression for the density matrix of the quantum state at arbitrary propagation distances and under arbitrary turbulence conditions.

However, the derivation of the IPE employs a Markov approximation to obtain the expression for the differential equation. In this approximation it is assumed that the medium is delta-correlated with itself along the propagation direction. For the derivation of the first-order differential equation of the IPE, one effectively assumes that the infinitesimal propagation step size is larger than the intrinsic scale, which in this case is the outer scale of the turbulence. To obtain the differential equation, one takes the limit where the step size goes to zero. On the other hand, the outer scale is assumed to go to infinity, allowing one to use the Kolmogorov turbulence model. This, seems to be a clear contradiction without a suitable justification. To some extent, the fact that the refractive index fluctuations are very small and thus allows light to propagate over long distances with minimal effect, mitigates this contradictory relationship between the step size and the outer scale. Still, our understanding of the evolution of photonic quantum states in turbulence would clearly benefit from a non-Markovian approach.

The Markov approximation is deeply ingrained in the work that has been done in the propagation of classical light through turbulence. Right from the start the assumption is made that the medium is delta-correlated along the propagation direction (see for instance \cite{scintbook}) and that for this reason the refractive index power spectral density can be treated as a two-dimensional function by setting the coordinate for the third dimension to zero. Thence, the theory is developed for all aspects of optical fields in turbulence, both within weak and strong fluctuation scenarios. Although the resulting theory seems to predict the behavior of classical light in turbulence adequately for the applications and conditions under consideration, one cannot currently say whether such a Markov approximation would be adequate for the evolution of quantum light in turbulence.

It is important to note that, although the system under investigation here deals with the evolution of a quantum state, it should not be confused with a non-Markovian quantum process. The latter concerns a situation where a system interacts with an environment such that the process needs to be described as an interacting quantum theory, formulated in terms of quantum mechanics. Such non-Markovian quantum processes are in general quite complex (see for instance \cite{petru}). In contrast, the non-Markovianity that one encounters in the evolution of a quantum state through turbulence is of a simpler nature. The process is linear --- there is no interaction --- and therefore it does not have a quantum bath that acts as the environment and interacts with the system. In the case of light propagating through turbulence, the effect of the medium is simply a continuous modulation process that extends over the propagation distance.

In this paper we consider a non-Markovian approach to study the evolution of photonic quantum states propagating through turbulence. We provide the derivation of a non-Markovian IPE, which takes the form of a second-order differential equation. The resulting equation has a form that does not in general have a solution. For this reason one needs to apply some simplifications or approximations to solve the differential equation. Here, we'll show two such approaches. In the first approach we assume that the turbulence is weak, which allows one to perform a perturbative expansion of the solution for the differential equation. The weak turbulence conditions can be considered as complimentary to the SPS model, which implies strong turbulence conditions \cite{notrunc}. The second approach is to modify the functional form of the differential equation. The resulting differential equation then does have a solution. Here, we'll only consider the single-photon case for this approach.

The paper is organized as follows. In Sec.~\ref{agter}, we provide a brief review of background material, followed by a discussion of the approach that we'll use to obtain a non-Markovian equation in Sec.~\ref{niemarkov}. The derivation of the non-Markovian IPE is shown in detail in Sec.~\ref{derive}. We provide the two different approaches to find solutions for the non-Markovian IPE in Secs.~\ref{oplos} and \ref{modi}, respectively. In Sec.~\ref{disc} we discuss some pertinent aspects of these solutions, followed by some conclusions in Sec.~\ref{concl}.

\section{Background}
\label{agter}

\subsection{Notation}

The discussions in this paper include both two-dimensional functions (such as the phase functions) and three-dimensional functions (such as the refractive index fluctuations). For this reason we need to define both two-dimensional and three-dimensional vectors to represent coordinate vectors. The two-dimensional coordinate vectors are always defined in the transverse plane, perpendicular to the propagation direction, the latter being the $z$-direction. For position coordinates, the two-dimensional position vector is denoted by a bold small ${\bf x}$, while the three-dimensional position vector is denoted by a bold capital ${\bf X}$. In the Fourier domain we prefer to work with spatial frequency vectors. The two-dimensional spatial frequency vector is denoted by a bold small ${\bf a}$, while the three-dimensional spatial frequency vector is denoted by a bold capital ${\bf A}$. Occasionally, we will also use the three-dimensional propagation vector, denoted by a bold capital ${\bf K}=2\pi{\bf A}$. The small $k$ is used to represent the wavenumber, which is not equal to $|{\bf K}|$.

During the analysis we'll obtain expressions for density matrices in terms of different sets of coordinates. Instead of denoting all these density matrices by the same symbol $\rho$, we rather avoid possible confusion by using different symbols $H$, $G$, etc. to represent the density matrices, depending on their arguments. We only use $\rho$ to represent the density matrix in generic discussions.

\subsection{Scintillation}
\label{spsagter}

For a thin enough slab of the turbulent medium, one can represent the scintillation process as a phase modulation. The phase functions that represent the turbulent medium in such a modulation process are random functions taken from an ensemble of such functions. Each one is obtained from an element of the ensemble of refractive index fluctuations $\delta n({\bf X})$, by an integration along the direction of propagation --- the $z$-direction. The phase functions are therefore defined by
\begin{equation}
\theta({\bf x}) = k \int_0^z \delta n({\bf X})\ {\rm d}z ,
\label{faseint}
\end{equation}
where $k$ is the wavenumber, given as $k=2\pi/\lambda$ in terms of the wavelength $\lambda$.

In the calculations of the evolution process, one often finds ensemble averages over phase functions, which give rise to the phase structure function in the following way
\begin{equation}
{\cal E}\!\left\{\exp\left[i\theta({\bf x}_1)-i\theta({\bf x}_2)\right]\right\} = \exp\left[-\frac{1}{2} D_{\theta} (\Delta {\bf x}) \right] ,
\label{expavg}
\end{equation}
where $\Delta {\bf x} = {\bf x}_1-{\bf x}_2$. Here
\begin{equation}
D_{\theta} (\Delta {\bf x}) = {\cal E}\!\left\{\right[\theta({\bf x}_1)-\theta({\bf x}_2)\left]^2\right\} ,
\label{strukt}
\end{equation}
is the phase structure function, which is related to the phase autocorrelation function 
\begin{equation}
D_{\theta} (\Delta {\bf x}) = 2 B_{\theta} \left(0\right) - 2 B_{\theta} (\Delta {\bf x}) ,
\label{dtnabt}
\end{equation}
The phase autocorrelation function is given by 
\begin{equation}
B_{\theta} (\Delta {\bf x}) = {\cal E}\!\left\{\theta({\bf x}_1)\theta({\bf x}_2)\right\} .
\label{kordeft}
\end{equation}
It is also referred to as a covariance function, because these random functions are assumed to have zero mean.

A similar relationship exists between the refractive index structure function and the refractive index autocorrelation function
\begin{equation}
D_n (\Delta {\bf X}) = 2 B_n \left(0\right) - 2 B_n (\Delta {\bf X}) ,
\label{dnnabn}
\end{equation}
where $\Delta {\bf X} = {\bf X}_1-{\bf X}_2$. The refractive index autocorrelation function is defined as
\begin{equation}
B_n (\Delta {\bf X}) = {\cal E}\!\left\{\delta n({\bf X}_1)\delta n({\bf X}_2)\right\} ,
\label{kordefn}
\end{equation}
and the refractive index structure function in the Kolmogorov theory is given by
\begin{equation}
D_n(\Delta {\bf X}) = C_n^2 |\Delta {\bf X}|^{2/3} .
\label{dndef}
\end{equation}

Using Eqs.~(\ref{faseint}) and (\ref{kordeft}), we express the two-dimensional phase autocorrelation function in terms of the three-dimensional refractive index autocorrelation function:
\begin{eqnarray}
B_{\theta} (\Delta {\bf x}) & = & k^2\!\int_0^z\!\!\int_0^z\!{\cal E}\!\left\{\delta n({\bf X}_1)\delta n({\bf X}_2)\right\} {\rm d}z_1 {\rm d}z_2 \nonumber \\ 
& = & k^2 \int_0^z \int_0^z B_n (\Delta {\bf X})\ {\rm d}z_1\ {\rm d}z_2 .
\label{kortnan}
\end{eqnarray}

Autocorrelation functions are related to power spectral density functions by the Wiener-Kinchine theorem \cite{statopt}. For the refractive index autocorrelation function we have 
\begin{equation}
B_n ({\bf X}) = \int \Phi_n({\bf K}) \exp(-i 2\pi {\bf A}\cdot{\bf X})\ {\rm d}^3 a ,
\label{wkn}
\end{equation}
where $\Phi_n({\bf K})$ is the refractive index power spectral density, which, in the Kolmogorov theory, reads \cite{scintbook}
\begin{equation}
\Phi_n ({\bf K}) = 0.033 (2\pi)^3 C_n^2 |{\bf K}|^{-11/3} ,
\label{klmgrv}
\end{equation}
where $C_n^2$ is the refractive index structure constant and the extra $(2\pi)^3$ factor is due to a difference in the definition of the Fourier transform \cite{iperr}. For the phase autocorrelation function we have 
\begin{equation}
B_{\theta} ({\bf x}) = \int \Phi_{\theta}({\bf a}) \exp(-i 2\pi {\bf a}\cdot{\bf x})\ {\rm d}^2 a ,
\label{wkt}
\end{equation}
where $\Phi_{\theta}({\bf a})$ is the phase power spectral density. Using Eqs.~(\ref{kortnan}), (\ref{wkn}) and (\ref{wkt}), one can express the phase autocorrelation function in terms of the refractive index power spectral density, which is given by
\begin{eqnarray}
\Phi_{\theta} ({\bf a}) & = & k^2 \int \int_{z_0}^{z} \int_{z_0}^{z} \exp[-i 2\pi (z_1-z_2) c] \nonumber \\
& & \times \Phi_n({\bf K})\ {\rm d} z_2\ {\rm d} z_1\ {\rm d} c .
\label{verwant}
\end{eqnarray}
The integrals over $z$ indicate that the refractive index fluctuations over the entire propagation path up to $z$ contribute to the behavior at $z$.

\subsection{Multiple phase screens}
\label{mpsagter}

The infinitesimal propagation principle, which allows a multiple-phase-screen approach, follows from considering the change in the photonic state due to an infinitesimal propagation through the medium. The operation of such an infinitesimal propagation on the density operator can be expressed by
\begin{equation}
\hat{\rho}(z) \rightarrow \hat{\rho}(z+\delta z) = dU \hat{\rho}(z) dU^{\dag} ,
\label{infquop}
\end{equation}
where $dU$ is a unitary operator representing the infinitesimal propagation through the turbulent medium. When the density operator is expressed as a density matrix in terms of some arbitrary discrete basis $\ket{m}$, the output density matrix elements, after the infinitesimal propagation, are given by
\begin{equation}
\rho_{mn}(z+\delta z) = \sum_{pq} \bra{m}dU\ket{p} \rho_{pq}(z) \bra{q}dU^{\dag}\ket{n} .
\label{infpsom}
\end{equation}

Using the paraxial wave equation in an inhomogeneous medium, given by \cite{scintbook}
\begin{equation}
\nabla_T^2 g({\bf X}) - i 2k\partial_z g({\bf X}) + 2k^2 \delta n({\bf X}) g({\bf X}) = 0 ,
\label{eomturb}
\end{equation}
where $g({\bf X})$ is the scalar electric field and $\delta n({\bf X})$ is the refractive index fluctuations, one can show that \cite{lindb}
\begin{equation}
\bra{m}dU\ket{p} = \delta_{mp} + i \delta z\ {\cal P}_{mp} + \delta z\ {\cal L}_{mp} ,
\label{infudef}
\end{equation}
where
\begin{equation}
{\cal P}_{mp}(z) \triangleq {2\pi^2 \over k} \int |{\bf a}|^2 G_m^*({\bf a},z) G_p({\bf a},z)\ {{\rm d}^2 a}
\label{kindef}
\end{equation}
and 
\begin{equation}
{\cal L}_{mp}(z) \triangleq - i k \iint G_m^*({\bf a},z) N({\bf a}-{\bf a}',z) G_p({\bf a}',z)\ {{\rm d}^2 a}\ {{\rm d}^2 a'} .
\label{dispdef}
\end{equation}
Here, $G_m({\bf a},z)$ and $N({\bf a},z)$ represent the two-dimensional transverse Fourier transforms of $g_m({\bf X})=\braket{x}{m}$ and $\delta n({\bf X})$, respectively.

The infinitesimal propagation of the density operator then leads to the following equation for each element in the ensemble \cite{lindb}
\begin{eqnarray}
\rho_{mn}(z_0+\delta z) & = & \rho_{mn}(z_0) + i \delta z \left[ {\cal P}, \rho(z_0) \right]_{mn} \nonumber \\
& & + \delta z \sum_p \left[ {\cal L}_{mp}(z_0) \rho_{pn}(z_0) \right. \nonumber \\
& & \left. + \rho_{mp}(z_0) {\cal L}_{pn}^{\dag}(z_0) \right] .
\label{lb1}
\end{eqnarray}
The right-hand side of Eq.~(\ref{lb1}) can be represented by an integral over a small range of $z$ to replace the factor of $\delta z$. If one were to compute the ensemble average of Eq.~(\ref{lb1}), the dissipative term (sum over $p$) would vanish, because ${\cal E}\!\{N\}=0$. One needs an expression with terms that are second-order in $N$ before computing the ensemble averages to have nonzero dissipative terms. The result of such ensemble averages would then contain autocorrelation functions of $N({\bf a},z)$.

\subsection{Markov approximation}
\label{markov}

The Markov approximation enters at the point where one computes the autocorrelation function of $N({\bf a},z)$
\begin{equation}
\Gamma_0({\bf a}_1,{\bf a}_2,z_1,z_2) = {\cal E}\!\{N({\bf a}_1,z_1) N^*({\bf a}_2,z_2)\} .
\label{nkor}
\end{equation}

One can model $N({\bf a},z)$ as
\begin{equation}
N({\bf a},z) = \int \left[ {\Phi_n({\bf K})\over\Delta^3} \right]^{1/2} \tilde{\chi}({\bf A}) \exp(-i 2\pi c z)\ {\rm d} c ,
\label{spek3d}
\end{equation}
where $\Delta$ is a correlation length in the frequency domain, $c$ is the `$z$-component' of ${\bf A}$ and $\tilde{\chi}({\bf A})$ is a normally distributed, delta-correlated, random complex function, with a zero mean. Hence,
\begin{equation}
{\cal E}\!\{\tilde{\chi}({\bf A}_1)\tilde{\chi}^*({\bf A}_2)\} = \Delta^3 \delta_3({\bf A}_1-{\bf A}_2) .
\label{deltkor}
\end{equation}
Since $\delta n({\bf X})$ is a real-valued function, the random complex function also obeys $\tilde{\chi}^*({\bf A})=\tilde{\chi}(-{\bf A})$. 

With the aid of Eq.~(\ref{spek3d}) we write Eq.~(\ref{nkor}) as
\begin{eqnarray}
\Gamma_0({\bf a}_1,{\bf a}_2,z_1,z_2) & = & \delta_2({\bf a}_1-{\bf a}_2) \int \exp[-i 2\pi (z_1-z_2) c_1] \nonumber \\
& & \times \Phi_n({\bf K}_1)\ {\rm d} c_1 .
\label{nkorc}
\end{eqnarray}

In the Markov approximation it is assumed that only the values of the field and the medium at $z$ contribute to the behavior at $z$. This assumption implies that the refractive index fluctuations are delta-correlated along the $z$-direction. The result is that one can substitute $k_z=0$ ($c=0$) in $\Phi_n({\bf K})$. Making this substitution and evaluating the integrals in Eq.~(\ref{verwant}), one arrives at a simpler relationship given by
\begin{equation}
\Phi_{\theta} ({\bf a}) = z k^2 \Phi_n(2\pi{\bf a},0) .
\label{verwantm}
\end{equation}
The simpler expression for $\Phi_{\theta} ({\bf a})$ can in turn be used to simplify the model for $N$:
\begin{equation}
N({\bf a},z) = \tilde{\chi}({\bf a}) \left[ \frac{\Phi_{\theta}({\bf a})}{\Delta^2} \right]^{1/2} ,
\label{spek2d}
\end{equation}
where $\tilde{\chi}({\bf a})$ is now a two-dimensional random function, but other than that has the same properties as $\tilde{\chi}({\bf A})$.

The Markov approximation is introduced into Eq.~(\ref{nkorc}) by setting $k_z=0$ in $\Phi_n({\bf K})$, which gives
\begin{equation}
\Gamma_1({\bf a}_1,{\bf a}_2,z_0,z) \approx \frac{\delta z}{2} \delta_2({\bf a}_1-{\bf a}_2) \Phi_n(2\pi{\bf a}_1,0) ,
\label{gam2}
\end{equation}
where $\delta z=z-z_0$. The factor of $\delta z$ leads to a first-order differential equation --- the Markovian IPE \cite{lindb}.

\section{Non-Markovian approach}
\label{niemarkov}

For the non-Markovian approach, we proceed without setting $k_z=0$ in $\Phi_n({\bf K})$. As a result, the integrals in Eq.~(\ref{verwant}) need to be evaluated by using an explicit expression for $\Phi_n({\bf K})$. On the other hand, one can exploit the fact that $\delta z$ is small for infinitesimal propagations. This allows one to expand Eq.~(\ref{nkorc}) up to leading order in $\delta z$. As a result we have
\begin{equation}
\Gamma_1({\bf a}_1,{\bf a}_2,z_0,z) \approx {\delta z^2\over 2} \delta_2 ({\bf a}_1-{\bf a}_2) \int \Phi_n({\bf K}_1)\ {\rm d} c_1 .
\label{gam3}
\end{equation}
The factor of $\delta z^2$ (instead of just $\delta z$) suggests that the non-Markovian equation could be a second-order differential equation.

In the derivation in Sec.~\ref{derive} and Appendix~\ref{eenaf}, we'll find that $z_1=z_2=z$. Thus, the correlation function in Eq.~(\ref{nkor}) or (\ref{nkorc}) becomes independent of $z$, so that
\begin{eqnarray}
\Gamma_0({\bf a}_1,{\bf a}_2) & = & {\cal E}\!\{N({\bf a}_1,z) N^*({\bf a}_2,z)\} \nonumber \\
& = & \delta_2({\bf a}_1-{\bf a}_2) \Phi_1({\bf a}_1) ,
\label{nmkor}
\end{eqnarray}
where
\begin{equation}
\Phi_1({\bf a}_1) \triangleq \int \Phi_n({\bf K}_1)\ {\rm d} c_1 . 
\label{phi1def}
\end{equation}

Master equations for non-Markovian systems (for example, the Nakajima-Zwanzig equation \cite{naka,zwanz}) in general have the form 
\begin{equation}
\partial_z \rho(z) = \int_{z_0}^{z} K(z,z') \rho(z')\ {\rm d} z ,
\label{gennm}
\end{equation}
where $K(z,z')$ is a super-operator that represents the memory in the system. Taking another derivative with respect to $z$ on both sides,
\begin{equation}
\partial_z^2 \rho(z) = K(z,z) \rho(z) + \int_{z_0}^{z} \left[ \partial_z K(z,z') \right] \rho(z')\ {\rm d} z ,
\label{gennm2}
\end{equation}
one finds that the right-hand side still contains an integral over $z$. It is therefore not in general possible to describe non-Markovian systems in terms of a pure second-order differential equation, having no integrals over $z$. If, however, $K(z,z')=K(z')$ in Eq.~(\ref{gennm}), one would obtain
\begin{equation}
\partial_z^2 \rho(z) = K(z) \rho(z) ,
\label{gennm3}
\end{equation}
which does not contain an integral over $z$.

In the particular case under consideration, it is possible to obtain a pure second-order differential equation, having no integrals over $z$. Consider Eq.~(\ref{lb1}), expressed as a first-order differential equation
\begin{eqnarray}
\partial_z \rho_{mn}(z) & = & i \left[ {\cal P}(z), \rho(z) \right]_{mn} + \sum_p \left[ {\cal L}_{mp}(z) \rho_{pn}(z) \right. \nonumber \\
& & \left. + \rho_{mp}(z) {\cal L}_{pn}^{\dag}(z) \right] .
\label{lb2}
\end{eqnarray}
If one differentiates Eq.~(\ref{lb2}) on both sides with respect to $z$, replaces the resulting first derivatives $\partial_z \rho(z)$ again by Eq.~(\ref{lb2}) and computes the ensemble average, by taking into account that ${\cal E}\!\{N\} = {\cal E}\!\{\partial_z N\} = 0$, one obtains a result that reads
\begin{eqnarray}
\partial_z^2 \rho_{mn}(z) & = & i \left[ \partial_z {\cal P}(z), \rho(z) \right]_{mn} - \left[ {\cal P}(z), \left[ {\cal P}(z), \rho(z) \right] \right]_{mn} \nonumber \\
& & + \sum_{p,q} {\cal E}\!\left\{ 2 {\cal L}_{mp}(z) \rho_{pq}(z) {\cal L}_{qn}^{\dag}(z) \right. \nonumber \\
& & - {\cal L}_{mp}^{\dag}(z) {\cal L}_{pq}(z) \rho_{qn}(z) \nonumber \\
& & \left. - \rho_{mp}(z) {\cal L}_{pq}^{\dag}(z) {\cal L}_{qn}(z) \right\} . 
\label{lb3}
\end{eqnarray}
Here we used the fact that the ${\cal L}$'s are anti-hermitian: ${\cal L}_{mn}^{\dag}=-{\cal L}_{mn}$. We note that Eq.~(\ref{lb3}) does not contain any integrals over $z$.

Using Eqs.~(\ref{dispdef}), (\ref{spek2d}) and (\ref{nmkor}), we compute the ensemble average over the ${\cal L}_{pq}$'s. The result is \cite{lindb}
\begin{eqnarray}
\Lambda_{mnpq} & \triangleq & {\cal E}\!\{{\cal L}_{mp}(z) {\cal L}_{qn}^{\dag}(z)\} \nonumber \\
& = & k^2 \iiint G_m^*({\bf a}_1+{\bf a}_2,z) G_p({\bf a}_2,z) G_q^*({\bf a}_3,z) \nonumber \\
& & \times G_n({\bf a}_3+{\bf a}_1,z) \Phi_1({\bf a}_1)\ {{\rm d}^2 a_1}\ {{\rm d}^2 a_2}\ {{\rm d}^2 a_3} \nonumber \\
& = & k^2 \int W_{mp}({\bf a},z) W_{nq}^*({\bf a},z) \Phi_1({\bf a})\ {{\rm d}^2 a} ,
\label{verll0}
\end{eqnarray}
where
\begin{equation}
W_{ab}({\bf a},z) \triangleq \int G_a^*({\bf a}'+{\bf a},z) G_b({\bf a}',z)\ {{\rm d}^2 a'} .
\label{wdef}
\end{equation}
When two of the indices on the ${\cal L}_{pq}$'s are contracted, one can use the orthogonality and completeness conditions of the modal basis to show that \cite{lindb}
\begin{equation}
\sum_p \Lambda_{mnpp} = \delta_{mn} k^2 \int \Phi_1({\bf a})\ {{\rm d}^2 a} \triangleq \delta_{mn} \Lambda_T .
\label{verll2}
\end{equation}
Substituting Eqs.~(\ref{verll0}) and (\ref{verll2}) into Eq.~(\ref{lb3}), we obtain
\begin{eqnarray}
\partial_z^2 \rho_{mn}(z) & = & i \left[ \partial_z {\cal P}(z), \rho(z) \right]_{mn} - \left[ {\cal P}(z), \left[ {\cal P}(z), \rho(z) \right] \right]_{mn} \nonumber \\
& & + 2 k^2 \int \sum_{p,q} W_{mp}({\bf a},z) \rho_{pq}(z) W_{qn}^{\dag}({\bf a},z) \nonumber \\
& & \times \Phi_1({\bf a})\ {{\rm d}^2 a} - 2 \Lambda_T \rho_{mn}(z) .
\label{lb4}
\end{eqnarray}
The result in Eq.~(\ref{lb4}) is a general expression for the non-Markovian IPE in an arbitrary discrete basis for a single photon propagating through turbulence.

Below, we'll repeat this derivation in detail, but we'll perform the derivation in the plane wave basis (Fourier domain), which is more beneficial for the purpose of finding solutions for the differential equation \cite{notrunc}.

\section{The non-Markovian IPE}
\label{derive}

In the transverse Fourier domain, the paraxial wave equation in an inhomogeneous medium is given by
\begin{equation}
\partial_z G({\bf a},z) = i\pi\lambda |{\bf a}|^2 G({\bf a},z) - i k N({\bf a},z) \star G({\bf a},z) ,
\label{transgft}
\end{equation}
where $\star$ represents convolution. The first term on the right-hand side of Eq.~(\ref{transgft}) represents free-space propagation and the second term produces distortions due to the effect of the medium. 

It is convenient to work in a `rotating' frame in which the free-space term is removed. This is done by using
\begin{equation}
G({\bf a},z) = F({\bf a},z) \exp\left(i\pi\lambda z |{\bf a}|^2\right) ,
\label{irrot}
\end{equation}
to convert the paraxial wave equation in Eq.~(\ref{transgft}) into
\begin{eqnarray}
\partial_z F({\bf a},z) & = & - i k \int N({\bf a}-{\bf u},z) F({\bf u},z) \nonumber \\
& & \times \exp\left[-i\pi\lambda z \left(|{\bf a}|^2-|{\bf u}|^2\right) \right]\ {\rm d}^2 u .
\label{transgft0}
\end{eqnarray}
To derive a non-Markovian IPE for a single-photon input state from Eq.~(\ref{transgft0}), we assume that the input is a single-photon pure state in the plane wave basis, given (in the rotating frame) by
\begin{equation}
R({\bf a}_1,{\bf a}_2,z) = F({\bf a}_1,z) F^*({\bf a}_2,z) .
\label{enkfot}
\end{equation}
The derivation of the non-Markovian IPE for the single-photon input state in Eq.~(\ref{enkfot}) is shown in Appendix~\ref{eenaf}. The result is
\begin{eqnarray}
\partial_z^2 R({\bf a}_1,{\bf a}_2,z) & = & 2 k^2 \int \left\{ R({\bf a}_1-{\bf u},{\bf a}_2-{\bf u},z) \right. \nonumber \\
& & \times \exp\left[-i 2\pi\lambda z \left({\bf a}_1-{\bf a}_2\right)\cdot{\bf u} \right] \nonumber \\
& & \left. - R({\bf a}_1,{\bf a}_2,z) \right\} \Phi_1({\bf u})\ {\rm d}^2 u .
\label{dveenfot}
\end{eqnarray}
Although it has an integral over the Fourier variables ${\bf u}$, the non-Markovian IPE is a second-order differential equation without any integrals over $z$. 

\begin{widetext}
The expression, equivalent to Eq.~(\ref{dveenfot}), for the two-photon states is given by
\begin{eqnarray}
\partial_z^2 R({\bf a}_1,{\bf a}_2,{\bf a}_3,{\bf a}_4,z) & = & 2 k^2 \int \left\{ R({\bf a}_1-{\bf u},{\bf a}_2-{\bf u},{\bf a}_3,{\bf a}_4,z) \exp\left[-i 2\pi\lambda z \left({\bf a}_1-{\bf a}_2\right)\cdot{\bf u} \right] \right. \nonumber \\
& & + R({\bf a}_1,{\bf a}_2,{\bf a}_3-{\bf u},{\bf a}_4-{\bf u},z) \exp\left[-i 2\pi\lambda z \left({\bf a}_3-{\bf a}_4\right)\cdot{\bf u} \right] \nonumber \\
& & \left. - 2 R({\bf a}_1,{\bf a}_2,{\bf a}_3,{\bf a}_4,z) \right\} \Phi_1({\bf u})\ {\rm d}^2 u .
\label{dvtweefot}
\end{eqnarray}
\end{widetext}

To aid the solution of the non-Markovian IPE we cast it in a form that decouples the $z$-dependence from the Fourier variables. This is done in a similar way as in \cite{notrunc}, by performing the following steps. 

First, we redefine the Fourier variables (spatial frequencies) in terms of sums and differences, defined by
\begin{eqnarray}
{\bf a}_1 & = & {\bf a}+\frac{1}{2}{\bf a}_d \\
{\bf a}_2 & = & {\bf a}-\frac{1}{2}{\bf a}_d .
\end{eqnarray}
The state is then also refined
\begin{eqnarray}
R({\bf a}_1,{\bf a}_2,z) & = & R({\bf a}+{\bf a}_d/2,{\bf a}-{\bf a}_d/2,z) \nonumber \\
& \triangleq & S({\bf a},{\bf a}_d,z) .
\end{eqnarray}
The expression in Eq.~(\ref{dveenfot}) then becomes
\begin{eqnarray}
\partial_z^2 S({\bf a},{\bf a}_d,z) & = & 2 k^2 \int \left[ S({\bf a}-{\bf u},{\bf a}_d,z) \exp\left(-i 2\pi\lambda z {\bf a}_d\cdot{\bf u} \right) \right. \nonumber \\
& & \left. - S({\bf a},{\bf a}_d,z) \right] \Phi_1({\bf u})\ {\rm d}^2 u .
\label{dv7}
\end{eqnarray}

The next step is to perform an inverse Fourier transform with respect to the sum coordinates:
\begin{equation}
H({\bf x},{\bf a}_d,z) = \int S({\bf a},{\bf a}_d,z) \exp(-i 2\pi {\bf a}\cdot {\bf x})\ {\rm d}^2 a .
\end{equation}
Equation~(\ref{dv7}) then reads
\begin{equation}
\partial_z^2 H({\bf x},{\bf a}_d,z) = - 2 k^2 Q(\lambda z {\bf a}_d+{\bf x}) H({\bf x},{\bf a}_d,z) ,
\label{dv9}
\end{equation}
where
\begin{equation}
Q({\bf x}) \triangleq \int \left[ 1-\exp(-i 2\pi {\bf x}\cdot{\bf u}) \right] \Phi_1({\bf u})\ {\rm d}^2 u  .
\label{qint}
\end{equation}

By combining the integral in Eq.~(\ref{qint}) with the definition in Eq.~(\ref{phi1def}), we find that $Q({\bf x})$ is related to the refractive index structure function, with $z=0$
\begin{equation}
Q({\bf x}) = \frac{1}{2} D_n ({\bf x},0) = \frac{1}{2} C_n^2 |{\bf x}|^{2/3} .
\label{qint2}
\end{equation}
Using the expression in Eq.~(\ref{qint2}), we obtain an expression for the single-photon non-Markovian IPE, given by 
\begin{equation}
\partial_z^2 H(z) = - k^2 C_n^2 H(z) |\lambda z {\bf a}_d+{\bf x}|^{2/3} .
\label{nmipe}
\end{equation}
where $H(z)\equiv H({\bf x},{\bf a}_d,z)$. The equivalent expression for the two-photon case is 
\begin{eqnarray}
\partial_z^2 H(z) & = & - k^2 C_n^2 H(z) \left( |\lambda z {\bf a}_d+{\bf x}_1|^{2/3} \right. \nonumber \\
& & \left. + |\lambda z {\bf b}_d+{\bf x}_2|^{2/3} \right) ,
\label{nm2ipe}
\end{eqnarray}
where $H(z)\equiv H({\bf x}_1,{\bf a}_d,{\bf x}_2,{\bf b}_d,z)$.

\section{Perturbative solution}
\label{oplos}

For the first method to solve the differential equation in Eq.~(\ref{nmipe}), we assume that the turbulence is weak enough [$C_n$ in Eq.~(\ref{nmipe}) is small enough] to allow a perturbative approach. This approach has the benefit that it can be generalized to the two-photon case, but first we'll consider the single-photon case.

\subsection{Single-photon state}

Consider a second-order differential equation given by Eq.~(\ref{gennm3}), but with the coupling constant $g$, which is proportional to the turbulence strength, made explicit
\begin{equation}
\partial_z^2 \rho(z) = g K(z) \rho(z) .
\label{gennm4}
\end{equation}
Expand the solution as an asymptotic series in $g$
\begin{equation}
\rho(z) = \rho_0(z) + g \rho_1(z) + g^2 \rho_2(z) + .... ,
\label{pertrho}
\end{equation}
and substitute it back into Eq.~(\ref{gennm4}). Setting g=0, one obtains the zeroth-order perturbation
\begin{equation}
\partial_z^2 \rho_0(z) = 0 .
\label{pert0}
\end{equation}
Its solution must satisfy the initial conditions. 

The two initial conditions for the second-order differential equation in Eq.~(\ref{gennm4}) can be stated as follows: 
\begin{enumerate}
\item the initial rate of change of the state is zero 
\begin{equation}
\left.\partial_z \rho(z) \right|_{z=0}=0 ,
\label{randv1}
\end{equation}
and 
\item the state at $z=0$ is given by the input state
\begin{equation}
\rho(0) = \rho_{\rm in} .
\label{randv2}
\end{equation}
\end{enumerate}

The solution of Eq.~(\ref{pert0}) that satisfies these initial conditions is
\begin{equation}
\rho_0(z) = \rho_{\rm in} .
\label{popl0}
\end{equation}

The first-order perturbation is obtained by taking a derivative with respect to $g$ before setting $g=0$. The resulting equation
\begin{equation}
\partial_z^2 \rho_1 (z) = K(z) \rho_0(z) = K(z) \rho_{\rm in} ,
\label{pert1}
\end{equation}
has a solution satisfying the initial conditions, given by
\begin{equation}
\rho_1(z) = \rho_{\rm in} \int_0^z \int_0^{z_2} K(z_1)\ {\rm d} z_1\ {\rm d} z_2  .
\label{popl1}
\end{equation}

Here we'll only go to sub-leading order in $g$. Therefore, our total solution, obtained from Eqs.~(\ref{popl0}) and (\ref{popl1}), is 
\begin{equation}
\rho(z) = \rho_{\rm in} \left[ 1+\int_0^z \int_0^{z_2} K(z_1)\ {\rm d} z_1\ {\rm d} z_2 \right] ,
\label{pertopl}
\end{equation}
where we reabsorbed $g$ into $K(z)$. 

To obtain an explicit expression for Eq.~(\ref{pertopl}), one needs to evaluate the double $z$-integration of $K(z)$. The expression for $K(z)$ for the single-photon case, according to Eq.~(\ref{nmipe}), is
\begin{equation}
K(z) = - k^2 C_n^2 \left(|\lambda z {\bf a}_d+{\bf x}|^2\right)^{1/3} .
\label{kdef}
\end{equation}
The solution in Eq.~(\ref{pertopl}) can thus be expressed as 
\begin{equation}
\rho(z) = \rho_{\rm in} \left[ 1 - k^2 C_n^2 \int_0^z \int_0^{z_2} P(z)^{1/3}\ {\rm d} z_1\ {\rm d} z_2 \right]
\label{pertopln}
\end{equation}
where 
\begin{equation}
P(z) = |\lambda z {\bf a}_d+{\bf x}|^2 =\lambda^2 z^2 |{\bf a}_d|^2 + 2\lambda z ({\bf a}_d\cdot{\bf x})+|{\bf x}|^2 .
\label{pol0}
\end{equation}

The evaluation of the integrations over $z$ in Eq.~(\ref{pertopln}) is briefly discussed in Appendix~\ref{strukint}. The result, in the $H$-notation of Eq.~(\ref{nmipe}), is given by
\begin{widetext}
\begin{eqnarray}
H({\bf x},{\bf a}_d,z) & = & H_{\rm in}({\bf x},{\bf a}_d) \left\{ 1 + \frac{3}{8} \frac{k^2 C_n^2 \left[\left(|{\bf a}_d|^2 \lambda z + {\bf a}_d\cdot{\bf x}\right)^2+\left({\bf a}_d\times{\bf x}\right)^2\right]^{4/3}-|{\bf a}_d|^{8/3} |{\bf x}|^{8/3}}{\lambda^2 |{\bf a}_d|^{14/3}} \right. \nonumber \\
& & + \frac{k^2 C_n^2 \left({\bf a}_d\cdot{\bf x}\right) \left(|{\bf a}_d|^2 \lambda z + {\bf a}_d\cdot{\bf x}\right) \left|{\bf a}_d\times{\bf x}\right|^{2/3}}{\lambda^2 |{\bf a}_d|^{14/3}} {_2{\rm F}_1} \left[\left(\frac{-1}{3},\frac{1}{2}\right),\left(\frac{3}{2}\right),\frac{-\left({\bf a}_d\cdot{\bf x}\right)^2}{\left({\bf a}_d\times{\bf x}\right)^2}\right] \nonumber \\
& & \left. - \frac{k^2 C_n^2 \left(|{\bf a}_d|^2 \lambda z + {\bf a}_d\cdot{\bf x}\right)^2 \left|{\bf a}_d\times{\bf x}\right|^{2/3}}{\lambda^2 |{\bf a}_d|^{14/3}} {_2{\rm F}_1} \left[\left(\frac{-1}{3},\frac{1}{2}\right),\left(\frac{3}{2}\right),\frac{-\left(|{\bf a}_d|^2 \lambda z + {\bf a}_d\cdot{\bf x}\right)^2}{\left({\bf a}_d\times{\bf x}\right)^2}\right] \right\} ,
\label{pertnaz}
\end{eqnarray}
\end{widetext}
where $H_{\rm in}({\bf x},{\bf a}_d)$ is the input state, ${_2{\rm F}_1}$ denotes a hyper-geometrical function and where we used the identity
\begin{equation}
({\bf A}\cdot{\bf B})^2+|{\bf A}\times{\bf B}|^2 = |{\bf A}|^2 |{\bf B}|^2 .
\label{vekid}
\end{equation}

The expression in Eq.~(\ref{pertnaz}) depends on a mixture of Fourier and position domain coordinates. It is preferable to obtain an expression that only depends on position domain coordinates. The expression in Eq.~(\ref{pertnaz}) has the form 
\begin{equation}
H({\bf x},{\bf a}_d,z) = H_{\rm in}({\bf x},{\bf a}_d) T({\bf x},{\bf a}_d,z) .
\label{genopl0}
\end{equation}
where the function $T(\cdot)$ is given by the part in curly brackets in Eq.~(\ref{pertnaz}). 

The general approach to obtain a position domain expression for the solution of the non-Markovian IPE, is to perform the steps of Sec.~\ref{derive} in reverse, keeping the expressions in terms of sum and difference coordinates all the way through. Starting from Eq.~(\ref{genopl0}), we first convert the expression completely to the Fourier domain by performing Fourier transforms with respect to ${\bf x}$. Then we add the free-space phase factor and perform an inverse Fourier transform on all coordinates to obtain the position space expression. Finally, one may simplify the expression by redefining the integration variables, using for instance ${\bf u}_s\rightarrow \lambda z {\bf a}$ and/or ${\bf u}_d\rightarrow {\bf x}_d-{\bf u}$. The resulting position domain expression reads
\begin{eqnarray}
G({\bf x}_s,{\bf x}_d,z) & = & \int G_{\rm in}({\bf x}_s-\lambda z {\bf a},{\bf u}) T \left({\bf u},\frac{{\bf x}_d-{\bf u}}{\lambda z},z\right) \nonumber \\ 
& & \times \exp \left[ -i 2\pi {\bf a}\cdot \left( {\bf x}_d-{\bf u} \right) \right]\ {\rm d}^2 a\ {\rm d}^2 u ,
\label{xopl}
\end{eqnarray}
in terms of the sums and differences of the position coordinates
\begin{eqnarray}
{\bf x}_s & = & \frac{1}{2} \left({\bf x}_1+{\bf x}_2\right) \\
{\bf x}_d & = & {\bf x}_1-{\bf x}_2 .
\end{eqnarray}
 
In Eq.~(\ref{xopl}), $T(\cdot)$, which is the same function that appears in Eq.~(\ref{genopl0}), serves as a kernel function for the propagation process. We'll refer to it as the (non-Markovian) turbulence propagation kernel.

Comparing the arguments of the turbulence propagation kernels in Eq.~(\ref{genopl0}) and (\ref{xopl}), one finds that the position domain expression requires the replacements ${\bf x}\rightarrow {\bf u}$ and ${\bf a}_d\rightarrow ({\bf x}_d-{\bf u})/\lambda z$. This leads to the following replacements for the quantities appearing in Eq.~(\ref{pertnaz}) 
\begin{eqnarray}
|{\bf x}| & \rightarrow & |{\bf u}| \nonumber \\ 
|{\bf a}_d| & \rightarrow & \frac{|{\bf x}_d-{\bf u}|}{\lambda z} \nonumber \\ 
({\bf a}_d\cdot{\bf x}) & \rightarrow & \frac{\left({\bf x}_d\cdot{\bf u}-|{\bf u}|^2\right)}{\lambda z} .
\label{verv0}
\end{eqnarray}

We use these replacements to obtain a position domain expression for the turbulence propagation kernel
\begin{widetext}
\begin{eqnarray}
T \left({\bf u},\frac{{\bf x}_d-{\bf u}}{\lambda z},z\right) & = & 1 + \frac{g t^2}{w_0^{2/3}|{\bf x}_d-{\bf u}|^{14/3}} \left\{ \frac{3}{8} \left( |{\bf x}_d|^{8/3}-|{\bf u}|^{8/3}\right) |{\bf x}_d-{\bf u}|^{8/3} \right. \nonumber \\
& & + \left({\bf x}_d\cdot{\bf u}-|{\bf u}|^2\right) \left(|{\bf x}_d|^2-{\bf x}_d\cdot{\bf u}\right) |{\bf x}_d\times{\bf u}|^{2/3} {_2{\rm F}_1} \left[\left(\frac{-1}{3},\frac{1}{2}\right),\left(\frac{3}{2}\right),\frac{-\left({\bf x}_d\cdot{\bf u}-|{\bf u}|^2\right)^2}{|{\bf x}_d\times{\bf u}|^2}\right] \nonumber \\
& & \left. - \left(|{\bf x}_d|^2-{\bf x}_d\cdot{\bf u}\right)^2 |{\bf x}_d\times{\bf u}|^{2/3} {_2{\rm F}_1} \left[\left(\frac{-1}{3},\frac{1}{2}\right),\left(\frac{3}{2}\right),\frac{-\left(|{\bf x}_d|^2-{\bf x}_d\cdot{\bf u}\right)^2}{|{\bf x}_d\times{\bf u}|^2}\right] \right\} ,
\label{kernel0}
\end{eqnarray}
\end{widetext}
where 
\begin{equation}
t \triangleq \frac{\lambda z}{\pi w_0^2} ,
\label{tnaz}
\end{equation}
is a normalized propagation distance and 
\begin{equation}
g \triangleq \frac{4{\cal T}}{\Theta^4} ,
\label{gdef}
\end{equation}
is a dimensionless coupling constant. The expression of this coupling constant is obtained by considering the complete expression Eq.~(\ref{pertopln}) in terms of dimensionless quantities. The details of this analysis is provided in Appendix \ref{koppel}. 

Notice that in Eq.~(\ref{kernel0}) the propagation distance only appears together with the coupling constant in front of the dissipative term and not anywhere inside the curly brackets. In fact, there are no dimension parameters inside the curly brackets, only the difference in position coordinates ${\bf x}_d={\bf x}_1-{\bf x}_2$ and the integration variables ${\bf u}$. However, some $z$ dependence also enters via the arguments of the input state in Eq.~(\ref{xopl}).

\subsection{Two photon state}

The result in Eq.~(\ref{kernel0}), together with Eq.~(\ref{xopl}), represents a perturbative solution for the single-photon differential equation given in Eq.~(\ref{nmipe}). One can generalize this solution to the two-photon case. The general perturbative solution for the two-photon case, analogous to Eq.~(\ref{pertopl}), is 
\begin{eqnarray}
\rho(z) & = & \rho_{\rm in} \left[ 1+\int_0^z \int_0^{z_2} K_1(z_1)\ {\rm d} z_1\ {\rm d} z_2 \right. \nonumber \\ 
& & \left. + \int_0^z \int_0^{z_2} K_2(z_1)\ {\rm d} z_1\ {\rm d} z_2 \right] ,
\label{pert2opl}
\end{eqnarray}
where $K_1(z)$ and $K_2(z)$ are associated with the two photons, respectively.

The single-photon turbulence propagation kernel, given in Eq.~(\ref{kernel0}), has the form
\begin{equation}
T \left({\bf u},\frac{{\bf x}_d-{\bf u}}{\lambda z},z\right) = 1 + g W({\bf u},{\bf x}_d,z) ,
\label{trubkern1}
\end{equation}
where $W(\cdot)$ is given by the part in Eq.~(\ref{kernel0}) that is multiplied by $g$. For the two-photon case, the form of the expression of the turbulence propagation kernel, simply becomes
\begin{eqnarray}
T_2 \left({\bf u}_1,{\bf x}_{1d},{\bf u}_2,{\bf x}_{2d},z\right)  & = & 1 + g W({\bf u}_1,{\bf x}_{1d},z) \nonumber \\ 
& & + g W({\bf u}_2,{\bf x}_{2d},z) ,
\label{trubkern2}
\end{eqnarray}
where $W(\cdot)$ is the same function as in Eq.~(\ref{trubkern1}).

\section{Modified differential equation}
\label{modi}

Here we consider an alternative approach to solve the differential equation in Eq.~(\ref{nmipe}). The idea is that, although the differential equation in Eq.~(\ref{gennm3}) does not have a solution in general, it does have solutions when $K(z)$ has a particular functional form. In what follows, we'll consider one such example.

The differential equation in Eq.~(\ref{nmipe}) can be written as
\begin{equation}
\partial_z^2 \rho(z) = - k^2 C_n^2 P(z)^{1/3} \rho(z) ,
\label{gennm5}
\end{equation}
where $P(z)$ is given in Eq.~(\ref{pol0}). With the aid of Eq.~(\ref{vekid}), one can express $P(z)$ as
\begin{equation}
P(z) = \frac{\left[z\lambda|{\bf a}_d|^2 + ({\bf a}_d\cdot{\bf x})\right]^2 + |{\bf a}_d\times{\bf x}|^2}{|{\bf a}_d|^2} .
\label{pol1}
\end{equation}

A special case that does allow a solution for Eq.~(\ref{nmipe}), is when the cross-product term in Eq.~(\ref{pol1}) is neglected. The differential equation then has the form
\begin{equation}
\partial_z^2 \rho(z) = - \alpha^2 (z+\zeta)^{2/3} \rho(z) ,
\label{dvf}
\end{equation}
where
\begin{eqnarray}
\alpha & = & \frac{2\pi |{\bf a}_d|^{1/3} \sqrt{C_n^2}}{\lambda^{2/3}} \label{alphadef} \\
\zeta & = & \frac{({\bf a}_d\cdot{\bf x})}{\lambda|{\bf a}_d|^2} . \label{z0def}
\end{eqnarray}
The differential equation in Eq.~(\ref{dvf}) has the solution,
\begin{eqnarray}
\rho(z) & = & C_1 \sqrt{z+\zeta}\ {\rm J}_{3/8}\left[\frac{3\alpha}{4} (z+\zeta)^{4/3} \right] \nonumber \\ 
& & + C_2 \sqrt{z+\zeta}\ {\rm Y}_{3/8}\left[\frac{3\alpha}{4} (z+\zeta)^{4/3} \right] ,
\label{genopl}
\end{eqnarray}
where ${\rm J}_{\nu}$ and ${\rm Y}_{\nu}$ are Bessel functions of the first and second kind, respectively, and $C_1$ and $C_2$ are constant to be determined by the initial conditions, given in Eqs.~(\ref{randv1}) and (\ref{randv2}).

Applying the first initial condition Eq.~(\ref{randv1}), one finds that the constants much have the forms
\begin{eqnarray}
C_1 & = & C_0 {\rm Y}_{-5/8}\left(\frac{3\alpha}{4} \zeta^{4/3} \right) \label{f1def} \\
C_2 & = & - C_0 {\rm J}_{-5/8}\left(\frac{3\alpha}{4} \zeta^{4/3} \right) , \label{f2def}
\end{eqnarray}
where $C_0$ is a constant, common to both $C_1$ and $C_2$. Substituting Eqs.~(\ref{f1def}) and (\ref{f2def}) into Eq.~(\ref{genopl}), one obtains an interim expression for the solution, given by
\begin{eqnarray}
\rho(z) & = & C_0 \sqrt{z+\zeta} \nonumber \\ 
& & \times \left\{ {\rm Y}_{-5/8}\left(\frac{3\alpha}{4} \zeta^{4/3} \right) {\rm J}_{3/8}\left[\frac{3\alpha}{4} (z+\zeta)^{4/3} \right] \right. \nonumber \\
& & \left. - {\rm J}_{-5/8}\left(\frac{3\alpha}{4} \zeta^{4/3} \right) {\rm Y}_{3/8}\left[\frac{3\alpha}{4} (z+\zeta)^{4/3} \right] \right\} . \nonumber \\
\label{opl1}
\end{eqnarray}

Now we apply the second initial condition Eq.~(\ref{randv2}) to the expression in Eq.~(\ref{opl1}) to obtain
\begin{eqnarray}
\rho(0) = \rho_{\rm in} & = & C_0 \sqrt{\zeta} \left[ {\rm Y}_{-5/8}\left(\frac{3\alpha}{4} \zeta^{4/3} \right) {\rm J}_{3/8}\left(\frac{3\alpha}{4} \zeta^{4/3} \right) \right. \nonumber \\ 
& & \left. - {\rm J}_{-5/8}\left(\frac{3\alpha}{4} \zeta^{4/3} \right) {\rm Y}_{3/8}\left(\frac{3\alpha}{4} \zeta^{4/3} \right) \right] \nonumber \\ 
& = & \frac{8 C_0}{3\pi \zeta^{5/6} \sqrt{\alpha}} ,
\label{randv3}
\end{eqnarray}
where we used the Wronskian 
\begin{equation}
{\rm J}_{\nu+1}(z) {\rm Y}_{\nu}(z)-{\rm Y}_{\nu+1}(z) {\rm J}_{\nu}(z) = \frac{2}{\pi z} ,
\label{wronsk}
\end{equation}
to obtain the last expression in Eq.~(\ref{randv3}). It gives a relationship between $C_0$ and $\rho_{\rm in}$, which is then used to replace $C_0$ in Eq.~(\ref{opl1}). The resulting solution reads
\begin{eqnarray}
\rho(z) & = & \frac{3\pi}{8} \rho_0 \zeta^{5/6} \sqrt{\alpha} \sqrt{z+\zeta} \nonumber \\ 
& & \times \left\{ {\rm Y}_{-5/8}\left(\frac{3\alpha}{4} \zeta^{4/3} \right) {\rm J}_{3/8}\left[\frac{3\alpha}{4} (z+\zeta)^{4/3} \right] \right. \nonumber \\
& & \left. - {\rm J}_{-5/8}\left(\frac{3\alpha}{4} \zeta^{4/3} \right) {\rm Y}_{3/8}\left[\frac{3\alpha}{4} (z+\zeta)^{4/3} \right] \right\} . \nonumber \\ 
\label{opl2}
\end{eqnarray}

The expression for the solution of the simplified differential equation that satisfies the initial conditions, is obtained from Eq.~(\ref{opl2}) by substituting Eqs.~(\ref{alphadef}) and (\ref{z0def}), into it. We express the result
\begin{eqnarray}
H({\bf x},{\bf a}_d,z) & = & \frac{\pi}{2} \beta H_{\rm in}({\bf x},{\bf a}_d) \left(z\lambda|{\bf a}_d|^2+{\bf a}_d\cdot{\bf x}\right)^{1/2} \nonumber \\
& & \times \frac{\left({\bf a}_d\cdot{\bf x}\right)^{5/6}}{|{\bf a}_d|^{7/3}}\left\{ {\rm Y}_{-5/8}\left[\frac{\beta \left({\bf a}_d\cdot{\bf x}\right)^{4/3}}{|{\bf a}_d|^{7/3}}\right] \right. \nonumber \\
& & \times {\rm J}_{3/8}\left[\frac{\beta \left(z\lambda|{\bf a}_d|^2+{\bf a}_d\cdot{\bf x}\right)^{4/3}}{|{\bf a}_d|^{7/3}} \right] \nonumber \\
& & - {\rm J}_{-5/8}\left[\frac{\beta \left({\bf a}_d\cdot{\bf x}\right)^{4/3}}{|{\bf a}_d|^{7/3}}\right] \nonumber \\
& & \left. \times {\rm Y}_{3/8}\left[\frac{\beta \left(z\lambda|{\bf a}_d|^2+{\bf a}_d\cdot{\bf x}\right)^{4/3}}{|{\bf a}_d|^{7/3}} \right] \right\} ,
\label{opl3}
\end{eqnarray}
in terms of the $H$-notation of Eq.~(\ref{nmipe}). Here $H_{\rm in}({\bf x},{\bf a}_d)$ is the input state and
\begin{equation}
\beta \triangleq \frac{3\pi\sqrt{C_n^2}}{2\lambda^2} = \frac{3\sqrt{g}}{4\pi w_0^{7/3}} .
\label{betadef}
\end{equation}

To convert the expression in Eq.~(\ref{opl3}) to the position domain, we use an approach that is similar to the one followed to obtain Eq.~(\ref{xopl}) from Eq.~(\ref{genopl0}). However, here we find it more convenient to perform a shift in the integration variables
\begin{equation}
{\bf u} \rightarrow {\bf x}_d-{\bf u} ,
\label{skuif}
\end{equation}
with the result that Eq.~(\ref{xopl}) becomes
\begin{eqnarray}
G({\bf x}_s,{\bf x}_d,z) & = & \int G_0({\bf x}_s-\lambda z {\bf a},{\bf x}_d-{\bf u}) T \left({\bf x}_d-{\bf u},\frac{{\bf u}}{\lambda z},z\right) \nonumber \\ 
& & \times \exp \left( i 2\pi {\bf a}\cdot {\bf u} \right)\ {\rm d}^2 a\ {\rm d}^2 u .
\label{xopl1}
\end{eqnarray}

The resulting replacements in the arguments of the propagation kernel $T(\cdot)$ in this case are ${\bf x}\rightarrow {\bf x}_d-{\bf u}$ and ${\bf a}_d\rightarrow {\bf u}/\lambda z$, leading to the following replacements
\begin{eqnarray}
|{\bf a}_d| & \rightarrow & \frac{|{\bf u}|}{\lambda z} \nonumber \\ 
({\bf a}_d\cdot{\bf x}) & \rightarrow & \frac{\left({\bf x}_d\cdot{\bf u}-|{\bf u}|^2\right)}{\lambda z} .
\label{verv1}
\end{eqnarray}
Applying these replacements to the expression in Eq.~(\ref{opl3}), we obtain the following position domain expression for the single-photon turbulence propagation kernel
\begin{eqnarray}
T \left({\bf x}_d-{\bf u},\frac{{\bf u}}{\lambda z},z\right) & = & \frac{\pi\lambda z\beta}{2} 
\frac{\left({\bf u}\cdot{\bf x}_d\right)^{1/2}}{|{\bf u}|^{7/3}} \left({\bf u}\cdot{\bf x}_d-|{\bf u}|^2\right)^{5/6} \nonumber \\
& & \times \left\{ {\rm Y}_{-5/8}\left[\frac{z\lambda\beta \left({\bf u}\cdot{\bf x}_d-|{\bf u}|^2\right)^{4/3}}{|{\bf u}|^{7/3}}\right] \right. \nonumber \\
& & \times {\rm J}_{3/8}\left[\frac{z\lambda\beta \left({\bf u}\cdot{\bf x}_d\right)^{4/3}}{|{\bf u}|^{7/3}} \right] \nonumber \\
& & - {\rm J}_{-5/8}\left[\frac{z\lambda\beta \left({\bf u}\cdot{\bf x}_d-|{\bf u}|^2\right)^{4/3}}{|{\bf u}|^{7/3}}\right] \nonumber \\
& & \left. \times {\rm Y}_{3/8}\left[\frac{z\lambda\beta \left({\bf u}\cdot{\bf x}_d\right)^{4/3}}{|{\bf u}|^{7/3}} \right] \right\} .
\label{kernel1}
\end{eqnarray}

Unfortunately, the result given in Eq.~(\ref{kernel1}) cannot be directly generalized to a two-photon case, as in the perturbative case above. The reason is that, when the simplification that was applied to Eq.~(\ref{nmipe}) to give Eq.~(\ref{dvf}), is applied to Eq.~(\ref{nm2ipe}), the resulting differential equation is not solvable.

\section{Discussion}
\label{disc}

Here we only consider the MPS approach. Although the SPS model appears to follow from a Markovian approach, the leading contribution in the non-Markovian approach gives the same expression for the SPS model. The conclusions that one can derive from the SPS model are therefore applicable regardless of whether one considers a Markovian or non-Markovian approach.

One of the pertinent aspects the SPS model is that it gives the behavior of the state in terms of a single dimensionless parameter ${\cal W}=w_0/r_0$, where $w_0$ is the optical beam waist radius and $r_0$ is the Fried parameter \cite{fried}. The relationship between the Rytov variance \cite{scintbook}, which quantifies scintillation strength, and ${\cal W}$ indicates that, for a constant ${\cal W}$, the scintillation strength increases with propagation distance. The SPS model is only valid under weak scintillation conditions. Therefore, it breaks down when the propagation distance becomes too large. In the context of the evolution of an entangled quantum state propagating through turbulence, one finds that the SPS model can only describe this evolution correctly for the entire duration of a nonzero entanglement, if the turbulence is strong enough to complete this evolution over a relatively short propagation distance. 

As a result, one can conclude that the SPS model provides a tool to study quantum state evolution under strong turbulence conditions \cite{notrunc}. What is needed then is another model that can provide a tool to study quantum state evolution under weak turbulence conditions. For the Markovian approach, such a tool was presented in the form of the Markovian IPE \cite{ipe,notrunc}. Here we provide such a tool for the non-Markovian approach, where we exploit the weakness of the turbulence to obtain a perturbative solution.

We also provide another solution for the non-Markovian IPE that does not assume weak turbulence. This is obtained by modifying the differential equation for the non-Markovian IPE. The resulting modified differential equation only works to the single-photon case. Its solution cannot be generalized to the two-photon case, because the simplification that is used does not render a readily solvable differential equation in the two-photon case. Nevertheless, it is not inconceivable that one may be able to find a simplification that can be applied to the two-photon differential equation which would allow solutions. The resulting expressions would in general be even more complex than those that we obtained here.

The modification that is applied assumes that the cross-product between particular coordinate vectors gives a vanishing contribution to final result. This assumption depends on the particular input optical field. For instance, if the input optical field is a Gaussian beam, then the expectation value of this cross-product is zero.

\section{Conclusions}
\label{concl}

The propagation of a photonic quantum state through a turbulent atmosphere is considered in terms of a non-Markovian approach. This is done in contrast to the existing Markovian methods that have been proposed before. We derive a non-Markovian IPE, which takes the form of a second-order differential equation with respect to the propagation distance. The non-Markovian IPE contains no integrations over the propagation distance. The form of this second-order differential equation does not allow immediate solutions. 

To solve the non-Markovian IPE, we follow two different approaches. The first is to assume the turbulence is weak enough to allow a perturbative analysis. This approach gives a solution that contains hyper-geometrical functions. Although we obtain the solution for the single-photon case, it can be generalized to the two-photon case. 

The second approach is to apply a particular simplification to the form of the differential equation. The resulting simplified differential equation can be solved to give a solution in terms of Bessel functions of fractional order. It only applies to the single-photon case.

\appendix

\section{Derivation of the single-photon non-Markovian IPE}
\label{eenaf}

Here we show in detail the derivation of the single-photon non-Markovian IPE.

Differentiating Eq.~(\ref{enkfot}) with respect to $z$, one obtains
\begin{eqnarray}
\partial_z R({\bf a}_1,{\bf a}_2,z) & = & \left[\partial_z F({\bf a}_1,z)\right] F^*({\bf a}_2,z) \nonumber \\
& & + F({\bf a}_1,z) \left[\partial_z F^*({\bf a}_2,z)\right] .
\label{dv0}
\end{eqnarray}
Substitution of Eq.~(\ref{transgft0}) into Eq.~(\ref{dv0}) then leads to
\begin{eqnarray}
\partial_z R({\bf a}_1,{\bf a}_2,z) & = & - i k \int N({\bf a}_1-{\bf u},z) F({\bf u},z) F^*({\bf a}_2,z) \nonumber \\
& & \times \exp\left[-i\pi\lambda z \left(|{\bf a}_1|^2-|{\bf u}|^2\right) \right]\ {\rm d}^2 u \nonumber \\ 
& & + i k \int N^*({\bf a}_2-{\bf u},z) F({\bf a}_1,z) F^*({\bf u},z) \nonumber \\
& & \times \exp\left[i\pi\lambda z \left(|{\bf a}_2|^2-|{\bf u}|^2\right) \right]\ {\rm d}^2 u . \nonumber \\
\label{dv1}
\end{eqnarray}

\begin{widetext}
A second derivative with respect to $z$ produces several terms on the right-hand side, but only those terms that contain derivatives of $F$ and $F^*$ will lead to terms that are second-order in $N$. All the other terms fall away when ensemble averaging is performed. Hence, retaining only those terms that will survive ensemble averaging, we obtain
\begin{eqnarray}
\partial_z^2 R({\bf a}_1,{\bf a}_2,z) & = & - i k \int N({\bf a}_1-{\bf u},z) \left[F^*({\bf a}_2,z) \partial_z F({\bf u},z) + F({\bf u},z) \partial_z F^*({\bf a}_2,z)\right] \exp\left[-i\pi\lambda z \left(|{\bf a}_1|^2-|{\bf u}|^2\right) \right] \nonumber \\ 
& & + N^*({\bf a}_2-{\bf u},z) \left[F^*({\bf u},z) \partial_z F({\bf a}_1,z) + F({\bf a}_1,z) \partial_z F^*({\bf u},z)\right] \exp\left[i\pi\lambda z \left(|{\bf a}_2|^2-|{\bf u}|^2\right) \right]\ {\rm d}^2 u 
\label{dv2}
\end{eqnarray}
After substituting Eq.~(\ref{transgft0}) and its complex conjugate into Eq.~(\ref{dv2}) for the second time, we have
\begin{eqnarray}
\partial_z^2 R({\bf a}_1,{\bf a}_2,z) & = & k^2 \int \left\{ 2 N({\bf a}_1-{\bf u},z) N^*({\bf a}_2-{\bf v},z) F({\bf u},z) F^*({\bf v},z) \exp\left[i\pi\lambda z \left(|{\bf a}_2|^2-|{\bf v}|^2-|{\bf a}_1|^2+|{\bf u}|^2\right) \right] \right. \nonumber \\ 
& & - N({\bf a}_1-{\bf u},z) N({\bf u}-{\bf v},z) F({\bf v},z) F^*({\bf a}_2,z) \exp \left[-i\pi\lambda z \left(|{\bf a}_1|^2-|{\bf v}|^2\right) \right] \nonumber \\
& & \left. - N^*({\bf a}_2-{\bf u},z) N^*({\bf u}-{\bf v},z) F({\bf a}_1,z) F^*({\bf v},z) \exp\left[i\pi\lambda z \left(|{\bf a}_2|^2-|{\bf v}|^2\right) \right] \right\}\ {\rm d}^2 u\ {\rm d}^2 v .
\label{dv4}
\end{eqnarray}
Next we evaluate the ensemble averages of Eq.~(\ref{transgft0}), using Eq.~(\ref{nmkor}), to obtain
\begin{eqnarray}
\partial_z^2 R({\bf a}_1,{\bf a}_2,z) & = & k^2 \int \left\{ 2 \delta_2({\bf a}_1-{\bf u}-{\bf a}_2+{\bf v}) \Phi_1({\bf a}_1-{\bf u}) F({\bf u},z) F^*({\bf v},z) \exp\left[i\pi\lambda z \left(|{\bf a}_2|^2-|{\bf v}|^2-|{\bf a}_1|^2+|{\bf u}|^2\right) \right] \right. \nonumber \\  
& & - \delta_2({\bf a}_1-{\bf v}) \Phi_1({\bf a}_1-{\bf u}) F({\bf v},z) F^*({\bf a}_2,z) \exp \left[-i\pi\lambda z \left(|{\bf a}_1|^2-|{\bf v}|^2\right) \right] \nonumber \\
& & \left. - \delta_2({\bf a}_2-{\bf v}) \Phi_1({\bf a}_2-{\bf u}) F({\bf a}_1,z) F^*({\bf v},z) \exp\left[i\pi\lambda z \left(|{\bf a}_2|^2-|{\bf v}|^2\right) \right] \right\}\ {\rm d}^2 u\ {\rm d}^2 v ,
\label{dv5}
\end{eqnarray}
where $\Phi_1({\bf a}_1)$ is defined in Eq.~(\ref{phi1def}). One can now evaluate the integrals over ${\bf v}$ to remove the Dirac-delta functions. After some simplification one then obtains the expression for the single-photon non-Markovian IPE given in Eq.~(\ref{dveenfot}).
\end{widetext}

\section{Integration of the structure function}
\label{strukint}

The $z$-integrations that need to be evaluated in Eq.~(\ref{pertopln}), can be expressed by
\begin{equation}
{\cal S} = \int_0^z \int_0^{z_2} \left(|\lambda z {\bf a}_d+{\bf x}|^2\right)^{1/3}\ {\rm d} z_1\ {\rm d} z_2 .
\end{equation}
One can evaluate this integral in different ways, leading to expressions that may appear different, but represent the same function. Here, we only show one such approach, where we use a Dirac-delta function to remove the quadratic polynomial from under the power of $1/3$. The expression becomes
\begin{eqnarray}
{\cal S} & = & \int \int \int_0^z \int_0^{z_2} \exp[i2\pi b_0 (q_0-|\lambda z {\bf a}_d+{\bf x}|^2)]\ {\rm d} z_1\ {\rm d} z_2  \nonumber \\  
& & \times q_0^{1/3}\ {\rm d} b_0\ {\rm d} q_0 .
\label{ddel0}
\end{eqnarray}
An integration over $b_0$ will turn the exponential into a Dirac-delta function. 

First, we evaluate the $z$-integrations, which leads to an expression that contains error-functions. The error-functions are replaced by auxiliary integrals
\begin{equation}
{\rm erf} (A) \rightarrow \frac{2 A}{\sqrt{\pi}} \int_0^1 \exp(-\xi^2 A^2)\ {\rm d}\xi .
\label{aux0}
\end{equation}

Considering the $b_0$-integrals of the resulting expression, one finds two types of integrals. One is of the form that would produce Dirac-delta functions which are then removed after the $q_0$-integration. The other is of the form 
\begin{equation}
\int \frac{\sin[ (q_0-U) b_0]}{b_0}\ {\rm d} b_0 = {\rm sign}(q_0-U) \pi .
\label{intb0}
\end{equation}
The sign-function separates the integration range of $q_0$ into two regions that add with opposite signs. Once both the $b_0$- and $q_0$-integrations are evaluated, one obtains
\begin{eqnarray}
{\cal S} & = & - \frac{3}{8} \frac{\left[|{\bf x}|^2+2\lambda z \left({\bf a}_d\cdot{\bf x}\right)+\lambda^2 z^2|{\bf a}_d|^2\right]^{4/3}-|{\bf x}|^{8/3}}{\lambda^2|{\bf a}_d|^2} \nonumber \\  
& & +\frac{(\lambda z |{\bf a}_d|^2+\left({\bf a}_d\cdot{\bf x}\right))^2 }{\lambda^2|{\bf a}_d|^{14/3}}  \nonumber \\  
& & \times \int_0^1 \left[ \left({\bf a}_d\times{\bf x}\right)^2 + \left(|{\bf a}_d|^2 \lambda z + {\bf a}_d\cdot{\bf x}\right)^2 \xi^2\right]^{1/3} {\rm d}\xi \nonumber \\  
& & -\frac{ \left({\bf a}_d\cdot{\bf x}\right) \left[\lambda z |{\bf a}_d|^2+\left({\bf a}_d\cdot{\bf x}\right)\right] }{\lambda^2|{\bf a}_d|^{14/3}} \nonumber \\  
& & \times \int_0^1 \left[ \left({\bf a}_d\times{\bf x}\right)^2 + \left({\bf a}_d\cdot{\bf x}\right)^2 \xi^2 \right]^{1/3} {\rm d}\xi .
\end{eqnarray}
The remaining $\xi$-integrals are of the form
\begin{eqnarray}
\int_0^1 \left(A^2+B^2\xi^2\right)^{1/3} {\rm d}\xi & = &{_2{\rm F}_1} \left[\left(\case{-1}{3},\frac{1}{2}\right),\left(\frac{3}{2}\right),\frac{-B^2}{A^2}\right]  \nonumber \\  
& & \times (A^2)^{1/3}.
\label{intxi}
\end{eqnarray}
After evaluating the $\xi$-integrals and replacing the result into Eq.~(\ref{pertopln}), we obtain Eq.~(\ref{pertnaz}).

\section{Coupling constant}
\label{koppel}

To find an expression for a dimensionless coupling constant we express Eq.~(\ref{pertopl}), together with Eq.~(\ref{pol0}), in terms of dimensionless coordinates and parameters. These are defined by normalizing the original coordinates with the aid of the characteristic dimension parameters
\begin{eqnarray}
{\bf f} & \triangleq & \pi w_0 {\bf a}_d \nonumber \\ 
{\bf r} & \triangleq & \frac{{\bf x}}{w_0} \nonumber \\ 
t & \triangleq & \frac{\lambda z}{\pi w_0^2} .
\label{nkoord}
\end{eqnarray}
Here, $w_0$ is the radius of the optical beam. In terms of these coordinates, the solution in Eq.~(\ref{pertopl}) can be expressed as 
\begin{equation}
\rho(t) = \rho_{\rm in} \left[ 1 - \frac{4{\cal T}}{\Theta^4} \int_0^t \int_0^{t_2} \left(|t {\bf f}+{\bf r}|^2\right)^{1/3}\ {\rm d} t_1\ {\rm d} t_2 \right]
\end{equation}
where 
\begin{equation}
{\cal T} \triangleq C_n^2 w_0^{2/3} ,
\label{tdef}
\end{equation}
is a normalized turbulence strength, and 
\begin{equation}
\Theta \triangleq \frac{\lambda}{\pi w_0} ,
\label{bhoek}
\end{equation}
is the Gaussian beam divergence angle. The dimensionless combination of the dimension parameters in front of the dissipative term gives us the expression for the effective coupling constant:
\begin{equation}
g \triangleq \frac{4{\cal T}}{\Theta^4} .
\end{equation}


\end{document}